\documentclass{mn2e}
\usepackage[dvipsnames,usenames]{color}
\usepackage{colortbl}
\usepackage{epsf,times,mathptmx,graphicx}
\usepackage{amsmath}
\usepackage{mathtools}
\let\bibhang\relax

\usepackage{natbib}

\newcommand{\etal}{{et al}\/.}
\begin{document}
\title[Non-uniform magnetic fields]{Synchrotron and inverse-Compton
  emission from radio galaxies with non-uniform magnetic field and
  electron distributions}
\author[M.J.\ Hardcastle]
{M.J.\ Hardcastle\\
School of Physics, Astronomy and Mathematics, University of
  Hertfordshire, College Lane, Hatfield AL10 9AB\\
}

\maketitle
\begin{abstract}
I investigate the effect of non-uniform magnetic fields in the
extended structures of radio galaxies on the observed synchrotron and
inverse-Compton emission. On the assumption of an isotropic field,
with a given power spectrum and a Gaussian distribution of the
Cartesian components of the magnetic field strength, I derive a simple
integral that can be used numerically to calculate the synchrotron
emissivity from any electron population. In the case of power-law
spectra, I show that it is possible to estimate the difference between
the synchrotron emissivity from a region with such a field and that
from the commonly assumed arrangement where $B$ is constant
everywhere, though fully tangled, and that this difference is small,
though it increases if the electron energy density scales with the
field. An aged electron spectrum in such a field produces a
characteristic curved synchrotron spectrum which differs significantly
from the classical Jaffe-Perola spectrum, and I discuss some effects
that this might have on standard spectral age fitting. Finally, I show
that inverse-Compton scattering of the cosmic microwave background is
only moderately affected by such a field structure, with the effects
becoming more important if the electrons follow the field.
Magnetic-field estimates in the literature from combined synchrotron
and inverse-Compton modelling will give reasonable
estimates of the mean magnetic field energy density if the field is
non-uniform but isotropic.
\end{abstract}


\section{Introduction}

In order to estimate physical conditions in the extended components of
radio galaxies (jets, hotspots and lobes) from the observed
synchrotron and inverse-Compton emission
\citep[e.g.][]{Hardcastle+04,Kataoka+Stawarz05,Croston+05-2}, some
simplifying assumptions must be made. In inverse-Compton modelling, in
particular, it is typically assumed that the electron number density
as a function of electron energy $E$ (hereafter $N(E)$) is spatially
uniform, and that the magnetic field strength in the region of
interest (e.g.\ a jet, lobe or hotspot) can be characterised by
a single value of $B$, where the magnetic field is assumed to be fully
tangled on scales much smaller than the scale of interest.

These assumptions are clearly not valid in detail. It has been known
for many years that the radio (synchrotron) emission from lobes and
plumes shows complex fine (`filamentary') structure on a range of
scales \citep[e.g.][]{Fomalont+89,Carilli+91,Swain+98,Laing+08}. The
existence of these structures proves that at least one of $B$ and
$N(E)$ is not spatially uniform within these structures. The fact that
the radio spectra of different regions within the source are not
identical (often attributed to `spectral ageing', which will be
discussed below in more detail) is generally taken to imply that
$N(E)$ varies with position [though see \cite{Katz-Stone+93} and
\cite{Blundell+Rawlings00} for 
alternative views]. Finally, in a very few sources, observations of
inverse-Compton emission in which the radio/X-ray ratio varies as a
function of position
\citep[e.g.][]{Isobe+02,Hardcastle+Croston05,Isobe+05,Goodger+08} also
imply that at least one of $N(E)$ and $B$ vary on large spatial
scales, and, in conjunction with multi-frequency radio observations,
can be interpreted as showing that both must vary
\citep{Hardcastle+Croston05}. Moreover, the generally smooth
appearance of inverse-Compton emission from lobes when observed at
high resolution \citep{Hardcastle+Croston05} implies that the
variations in physical conditions responsible for the filamentary
radio structure are very unlikely to arise in $N(E)$ alone. Modelling
of the synchrotron and inverse-Compton emission from lobes should
therefore ideally take into account these variations in physical
conditions. In the present paper I shall focus on two aspects of this
modelling where non-uniform $B$ and $N(E)$ are likely to be important:
spectral ageing and inverse-Compton modelling.

`Spectral ageing' is the name given to the long-established technique
of using changes in the observed synchrotron spectrum to infer the
time since the particles in the region of interest were last
accelerated, and thus make inferences about the source age and/or the
lobe dynamics
\citep[e.g.][]{Jaffe+Perola73,Burch77,Myers+Spangler85,
    Alexander+Leahy87,Carilli+91,Liu+92,Mack+98,Murgia+99,Jamrozy+08}.
If $N(E)$ is initially a power law in energy as a result of the
acceleration process, then at some later time $t>0$ we expect $N(E)$
to have a characteristic form depending on $t$. Since for synchrotron
radiation the characteristic timescale for energy losses goes as
$1/E$, $N(E)$ at $t>0$ will always be characterized by depletion of
the highest-energy electrons. The appropriate form of $N(E)$ depends
on whether there is no pitch angle scattering of the electrons
\citep[hereafter KP]{Kardashev62,Pacholczyk70} or effective pitch
angle scattering \citep[hereafter JP]{Jaffe+Perola73}. In the KP
model, some very high-energy electrons survive indefinitely because
their pitch angles are very small and their energy losses negligible;
thus the high-energy cutoff of the electron spectrum is that imposed
by the original particle acceleration process, but there is a break in
the electron energy spectrum when integrated over pitch angle at the
point at which losses become non-negligible for electrons with larger
pitch angles. Pitch angle scattering is more plausible a priori
  for realistic, turbulent magnetic field configurations
  because of the expected resonant scattering of particles on 
  the field, which is effective on scales comparable to the Larmor
  radius of the electrons, and hence much smaller than the resolution
  of observations (see also discussion by \citealt{Carilli+91}). In
  the JP model, in which pitch angle scattering takes place, there is a
critical energy, a function of $t$, above which there are no remaining
electrons, and this in turn gives rise to an exponential cutoff in the
synchrotron spectrum at high energies.

Both JP and KP models are in principle models of the shape (not
normalization) of $N(E)$ and so are independent of our assumptions
about electron number densities and magnetic field strengths, but in
practice `JP spectra' and `KP spectra' are normally calculated and
fitted to data on the assumption of a uniform magnetic field strength,
and, moreover, it is often assumed that the field in which the
particle energy spectrum has evolved is the same as that in which it
is currently radiating. Early and important work on the problems with
these assumptions was done by \cite{Tribble93}, who showed numerically
that a distribution of magnetic fields would give rise to synchrotron
spectra that were significantly modified with respect to the JP and KP
spectra, a point reiterated for different assumptions about the
  distribution of field strengths by \cite{Eilek+Arendt96}. More
recent work \citep[e.g][]{Eilek+97,Kaiser05} has focused on analytical
modelling of two-field systems, where ageing can take place in both
fields but the emission is dominated by the high-field region;
however, while this is analytically tractable and suffices to show
that even simple variations of the field strength will give rise to
observable effects on the synchrotron spectrum, the assumption of only
two field strengths is a limitation of these models. With the
new-generation radio telescopes providing much better frequency
coverage of the synchrotron spectrum, it is timely to look again at
modifications to the spectral ageing model.

The inferences drawn from inverse-Compton emission from lobes are also
affected by inhomogeneous $N(E)$ and $B$. $N(E)$ of course affects
inverse-Compton emission directly. For the simplest case, that of
inverse-Compton scattering of the cosmic microwave background (CMB),
the inverse-Compton emissivity depends only on the number density of
electrons with $\gamma \approx 1000$, and since these are too low in
energy to be depleted by loss processes in typical radio source
lifetimes, the inverse-Compton emissivity simply depends on the
low-energy normalization of $N(E)$. However, inverse-Compton emission
is used to estimate $B$, by comparing the synchrotron emissivity,
which depends on $B$, with the inverse-Compton emissivity, which does
not: these estimates of a characteristic $B$ may be very wrong in the
presence of an inhomogeneous field. Synchrotron self-Compton emission
obviously depends on the distribution of $N(E)$ and $B$ in a much more
complicated way.

There are two possible ways to improve our understanding of the
expected synchrotron and inverse-Compton properties of radio galaxies.
In principle the best approach would be to carry out full numerical
modelling of the lobes, including all relevant physical processes that
could affect the electron energy spectrum. This approach was
pioneered, though with very low resolution and for axisymmetric jets,
by \cite{Matthews+Scheuer90a,Matthews+Scheuer90b}. More recently,
\cite{Tregillis+01} have carried out 3D MHD simulations including
acceleration processes, and the state of the art for this type of work
is represented by the type of code described by \cite*{Mendygral+12},
which can in principle give arbitrarily detailed inverse-Compton and
synchrotron spectra as a function of position. The key advantage of
such work is that the {\it relationship} between the particle and
field populations should be realistic, with no simplifying assumptions
necessary. The disadvantages are that such studies are computationally
very expensive, their results depend on the input assumptions about
the microphysics of, for example, particle acceleration and
dissipation of magnetic fields, and they necessarily only sample a
very small area of the parameter space that it is possible for
radio-loud AGN to occupy. As a result, while such work is essential to
provide a detailed understanding of these systems, it is difficult to
abstract from it general methods for interpreting observational data.

In the present paper I take the complementary approach of using
  analytical calculations, supplemented by simple time-independent
  numerical modelling. I revisit some of the analytical and numerical
calculations originally carried out by \cite{Tribble91},
\cite{Tribble93} and \cite{Tribble94}, giving a clear statement of the
analytic approach used and extending the numerical work using
currently available computing power. In Section 2 I describe the
general approaches used in the paper and discuss their applicability
to real radio sources. In Section 3 I apply them to observations of
aged synchrotron emission, noting in particular the important effect
of structure on finite-sized regions on the observed spectral shapes,
and in Section 4 I consider the consequences of the emission models
developed in Sections 2 and 3 for inverse-Compton analyses. Discussion
and prospects for further work are presented in Section 5.

\section{The methods}
\label{methods}

The analytical method used to describe synchrotron emission in a
random magnetic field was described by \cite{Tribble91}. The basic
assumption he used was that the field is a Gaussian random field,
i.e.\ that each component of the field at each point is drawn from a
Gaussian distribution with mean zero and identical dispersion; this
would be the case, for example, if the magnetic field structure is generated
by homogeneous, isotropic turbulence\footnote{In general we expect MHD
  turbulence to be anisotropic \citep[e.g.][]{Cho+Lazarian03}. My
  approach in this paper is to ignore anisotropy in the hope that this
  is a reasonable assumption for small enough regions of the source.
  Simulations of realistic MHD turbulence are beyond the scope of this
  paper.}. In this case, the distribution
of the magnitude of the magnetic field strength vector is a
Maxwell-Boltzmann distribution. We can then consider the expected
synchrotron emissivity for an arbitrary electron energy distribution.
Let the electron energy spectrum be described by $N(E)$, where $E$ is
the energy of the electron and $N(E){\rm d}E$ is the number density of
electrons with energies between $E$ and $E + {\rm d}E$. The
single-electron emissivity as a function of frequency is given by
\citep{Longair10} 
\begin{equation}
j(\nu) = \frac{\sqrt{3} Be^3 \sin \alpha}{8\pi^2 \epsilon_0 c
m_e}F(x) 
\end{equation}
where $B$ is the local magnitude of the magnetic field strength, $e$
is the charge on the electron, $\epsilon_0$ is the permittivity of
free space, $F(x)$ is defined \citep{Rybicki+Lightman79} as
\begin{equation}
F(x) = x \int^\infty_x K_{5/3}(z) {\rm d}z
\end{equation}
with $K_{5/3}$ the modified Bessel function of order $5/3$, and $x$ is
a dimensionless function of the frequency, field strength and energy:
\begin{equation}
x = \frac{\nu}{\nu_c} = \frac{4\pi m_e^3c^4 \nu}{3eE^2 B \sin\alpha}
\label{nu-crit}
\end{equation}
If we further assume that the pitch angle distribution is known (the
standard assumption being that the electron population is isotropic,
i.e. $p_\alpha = \frac{1}{2} \sin \alpha$), then we can write down an
integral that gives the emissivity from the entire population at a
given frequency:
\begin{equation}
J(\nu) = \int_0^\infty \int_0^\pi \int_{E_{\rm min}}^{E_{\rm max}} \frac{\sqrt{3} Be^3 \sin \alpha}{8\pi^2 \epsilon_0 c
m_e} F(x) N(E) p_\alpha p_B {\rm d}E\, {\rm d}\alpha\, {\rm d}B
\label{emissivity}
\end{equation}
(cf.\ \citealt{Eilek+Arendt96}) where $p_\alpha$ and $p_B$ are the
probability distributions of the pitch angle $\alpha$ and the magnetic
field strength $B$ respectively. As noted above, $p_B$ for a
Gaussian random field is the Maxwell-Boltzmann distribution, which we can write
in a general form with a parameter $a$:
\begin{equation}
p_B = \sqrt{\frac{2}{\pi}} \frac{B^2 \exp(-B^2/2a^2)}{a^3}
\end{equation}
It is convenient to consider a field with a given mean magnetic field
energy density, for which
\begin{equation}
\int B^2 p_B {\rm d}B = B_0^2
\end{equation}
which sets the value of the parameter $a = B_0/\sqrt{3}$.

Equation \ref{emissivity} can then be integrated -- in the general
case, numerically -- over the known electron, pitch angle and magnetic
field distributions to give a synchrotron spectrum. This is the method
used by \cite{Tribble91} and, although not explicitly stated,
\cite{Tribble93} must have used numerical integration of a version of
equation \ref{emissivity}.

\begin{figure}
\epsfxsize 8.5cm
\epsfbox{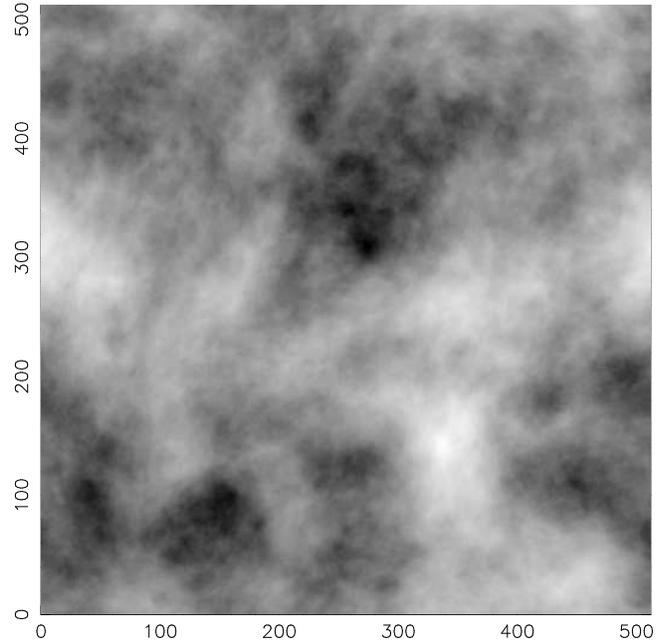}
\caption{Synchrotron surface brightness from a spatially uniform ($s=0$) power-law electron
  energy distribution and a spatially varying field as described in
  the text. Magnetic field with $\zeta = 5.667$, $k_{\rm min}=1$ and
  $k_{\rm max}=256$ is
  generated in a $512 \times 512 \times 512$ box and the synchrotron
  emissivity projected along the $z$-axis. Lighter shades of grey
  indicate brighter regions; the scale is logarithmic with 0.89 dex
  between the brightest and faintest regions.}
\label{synch-box}
\end{figure}

\cite{Tribble94} described a numerical method, which would be expected
to give the same results but which allows visualization of the
synchrotron-emitting region rather than merely a calculation of its
total emissivity, in which a box of random magnetic field is generated
and the emissivity of each cell can be calculated. A general method
for generating a random magnetic field with particular statistical
properties was described by \cite{Tribble91}. The same methods were
used by \cite{Murgia+04}, hereafter M04, modified to generate a
magnetic field that might be appropriate for a turbulent cascade, and
I closely follow the approach of M04 in what follows. Specifically, I
generate the three components of the Fourier transform of the magnetic
vector potential $\mathbf{\tilde{A}}(k)$ by drawing their complex
phases from a uniform distribution and their magnitudes from a
Rayleigh distribution whose controlling parameter $\sigma$ ($|A_k|$ in
the notation of Murgia et al) depends on the wavenumber $k$ as $\sigma
= k^{-\zeta}$ between minimum and maximum wavenumbers $k_{\rm min}$
and $k_{\rm max}$ and is taken to be zero elsewhere.(In the
  numerical realizations that follow
I take $k_{\rm min} = 1$ and $k_{\rm max}$ to be the maximum value
possible for the computational box used, so that the largest possible
range of spatial scales is present in the models.) The Fourier
transform of the magnetic field is then
\[
\mathbf{\tilde{B}}(k) = i\mathbf{k} \times \mathbf{\tilde{A}}(k)
\]
and I take the inverse three-dimensional Fourier transform of the
three components of this, using code provided by the FFTW library, to
find the real values of the components of the magnetic field vector
$\mathbf B$, which are then renormalized to physical units. (For most
of the
runs used in this paper this is done using a $512 \times 512 \times
512$ box.) This process, of course, gives rise to a Gaussian random
field with zero mean, and the power spectrum $|B_k|^2 \propto k^{-(\zeta
  -2)}$ (M04).

To compute the synchrotron emissivity from the box I use the method
suggested by eq. 22 of M04: that is, the effective magnetic field is
taken to be the component perpendicular to the line of sight (which I
take to be the $z$-direction of my Cartesian co-ordinates), so that
$B_\perp = |\mathbf{B}| \sin \theta = B \sin \theta$. Since only
electrons with pitch angles $\alpha = \theta$ contribute
non-negligibly to the resulting radiation, this means that the
single-electron critical frequency is everywhere proportional to
$B_\perp E^2$. For speed, rather than numerically integrating over the
electron energy density at every point in the grid, I compute the
emissivity for a single value of $B_\perp$ for all required
frequencies (using equation \ref{emissivity} but with $p_B$ and
$p_\alpha$ as delta functions), fit a cubic spline to it, and then
scale its normalization and base frequency using the values of
$B_\perp$ and $B$ appropriate to each cell.

An example of the projected synchrotron emissivity for such a magnetic
field configuration (assuming a power-law electron energy spectrum
with $N(E) \propto E^{-p}$: see below) is shown in Fig.
\ref{synch-box}; here we assume a uniform electron density throughout
the box. The appearance of the synchrotron emission from the region is
clearly expected to be strongly dependent on the power spectrum used
for the vector potential, i.e. on $k_{\rm min}$, $k_{\rm max}$ and
$\zeta$; it will also depend on the electron energy spectrum. One way
of characterizing the synchrotron emission is to find its power
spectrum: \cite{Eilek89} showed that a given power spectrum of the
magnetic field translates into a particular two-dimensional power
spectrum of the observed emission. Taking the two-dimensional
  Fourier transform of the projected emission and averaging its
  amplitude in radial bins to find the power on each spatial scale, I
find that, as expected, a power law is recovered, with the amplitude
going as $k^{-n}$ where $n \approx \frac{2}{3}(\zeta - 3)$ for flat
electron energy spectra ($p \approx 2$); there is only a very
weak dependence on the electron energy index $p$, in the sense that
higher values of $p$ give rise to higher values of $n$. Carrying
  out the same procedure on some high-quality radio images, I find
that power laws for the amplitudes of the observed synchrotron
emission from well-resolved bright radio lobes have $n \approx 2$ in
the range in which they are well described by a power law in $k$
(which in practice is a range of scales starting significantly
  below the size scale of the lobes and ending significantly above the
  nominal spatial resolution), so that we should adopt $\zeta \approx
6$ to reproduce these. In what follows I take a standard value of
$\zeta =5.667$, as used in the image shown in Fig. \ref{synch-box},
since this corresponds to a Kolmogorov spectrum for $B$. The fact that
the observed properties of radio lobes are consistent with a simple
model of magnetic turbulence together with a uniform electron energy
spectrum gives us some confidence that the model used here is at least
somewhat relevant to real systems.

What is of course missing from the numerically modelled
synchrotron-emitting regions is any of the large-scale filamentary
structure seen in real radio lobes. There are two reasons for this;
firstly, the simulated region cannot, because of the way it is
constructed, have any contribution from structures larger than the
half-size of the box: secondly, the most striking of the filamentary
structures are almost certainly a result of the large-scale internal
dynamics of the lobes, and are a symptom of the processes that drive
the turbulence rather than of the turbulence itself. I do not regard this
as a serious problem since the only modelling I intend to carry out
will be of regions small enough that the electron energy spectrum does
not vary significantly across them; that is, the models should be
taken to represent regions much smaller than the characteristic size
of the lobes (with the caveat that, of course, we have few
  observational constraints on the properties of magnetic turbulence
  on those scales).

\section{Synchrotron spectra}

\subsection{Power-law electron spectra}
\label{powerlaw}

The synchrotron spectrum of a region containing a non-uniform magnetic
field, for a given electron energy distribution, can now be
investigated. I choose to restrict this to the situation where the
electron energy spectrum is the same, apart from normalization, at all
points in space: this is reasonable if the diffusion time for
electrons throughout the region is much shorter than the loss time, so
that, averaging over loss times, all electrons probe the same mean
magnetic field strength. (See \cite{Tribble93} and \cite{Eilek+97} for
modelling in which this is not the case.) Since there is some evidence
that $N(E)$ does vary on the largest scales, these models are not
applicable on those scales.

We can begin by considering a power-law distribution of electrons,
$N(E) = N_0E^{-p}$. In this situation it is clear from the standard
arguments that we still expect a power-law synchrotron spectrum with
the usual frequency dependence, $J(\nu) \propto \nu^{-(p-1)/2}$. Only
the normalization can change. In fact, for the case where we have a
uniform $N(E)$ independent of $B$, we can see that if the mean
squared magnetic field strength is the same, $B^2_0$, the {\it total} electron
energy loss rate to synchrotron emission in the region must be the
same, and as the spectral shapes are the same except at the end points
of the power law, the normalizations should be essentially identical.
Another way of seeing this is to integrate the standard power-law
emissivity approximation,
\begin{equation}
J(\nu) = K N_0 \nu^{-{{(p-1)}\over 2}} B^{{(p+1)}\over 2}
\label{emiss-approx}
\end{equation}
where $K$ is a constant, over the distribution of magnetic field strengths, obtaining
\begin{equation}
\int J(\nu) p_B\, {\rm d}B =
\frac{2}{\sqrt{\pi}}\left(\frac{2}{3}\right)^\frac{p+1}{4} \Gamma\left(\frac{p+7}{4}\right) K N_0
\nu^{-\frac{(p-1)}{2}} B_0^{\frac{(p+1)}{2}}
\label{normchange}
\end{equation}
where the leading numerical term differs from unity by only a few per
cent over a physically reasonable range for $p$ (2--3); that is, the
emissivity from a Maxwell-Boltzmann distribution of field strengths is
only very slightly different from that from $p_B = \delta(B-B_0)$, a
fully tangled field with a constant magnetic field strength.

Somewhat more interesting is the case where the normalization of the
electron energy spectrum ($N_0$) is not a constant in space. It is clear that
if the variations in $N_0$ are uncorrelated with those in $B$, then
there will be no effect on the synchrotron emissivity integrated over
a sufficiently large volume -- we simply replace the constant $N_0$ in
the equations above with its mean value $\bar N_0$. More interesting
is the case where the local normalization depends on the magnetic field
strength. We may assume some general power-law relation,
\begin{equation}
N_0 \propto \left(\frac{B}{B_0}\right)^s
\label{s-definition}
\end{equation}
where $s=0$ corresponds to no dependence on $B$ and $s=2$ to what we
could think of as `local equipartition' -- the energy density of the
particles scales everywhere with that in the field. \cite{Eilek89}
suggests that intermediate values of $s$, 1--1.5, might be reasonable in
trans-sonic turbulence. Then the ratio between the emissivity in this
situation and that in the fully-tangled, uniform-density,
uniform-field case with the same mean magnetic field and particle
energy density becomes
\begin{equation}
\frac{\displaystyle \int B^s B^{{(p+1)}\over 2} p_B\,{\rm
    d}B}{\displaystyle B_0^{{(p+1)}\over 2}
  \int B^s p_B\,{\rm d}B} = \left(\frac{2}{3}\right)^\frac{p+1}{4}
\frac{\Gamma\left(\frac{s}{2} +
  \frac{p+7}{4}\right)}{\Gamma\left(\frac{s+3}{2}\right)}
\end{equation}
This is an almost linear function of $s$ which goes from $\sim 1$ at
$s=0$ to $\sim 1.5$ at $s=2$, more or less independent of $p$ in the
range discussed above. Thus we expect the total emissivity in models in
which the electron density follows the field strength to be up to a
factor 1.5 higher than in the uniform-density, constant-field-strength
models for a power-law spectrum. We also expect the power spectrum of
the resolved emissivity to have a steeper power-law index, since the effect of increasing
$s$ is to increase the contrast between the brightest and faintest
regions of the map. In practice this is a small effect, with the value
of $n$ increasing by only $\sim 0.1$ for $s=2$, $p=2.2$.

Finally, it is important to note that the preceding analysis is for
the idealized case where we integrate over an infinite volume. In real
systems, and in particular in systems where a turbulent magnetic field
power spectrum extends to the largest available scales, as discussed
in Section \ref{methods}, there will be considerable scatter in the
relationship between mean electron and magnetic field energy density
on the one hand, and emissivity on the other. We return to this point
below.

\subsection{Aged spectra}
\label{aged}

\begin{figure*}
\epsfxsize 8.7cm
\epsfbox{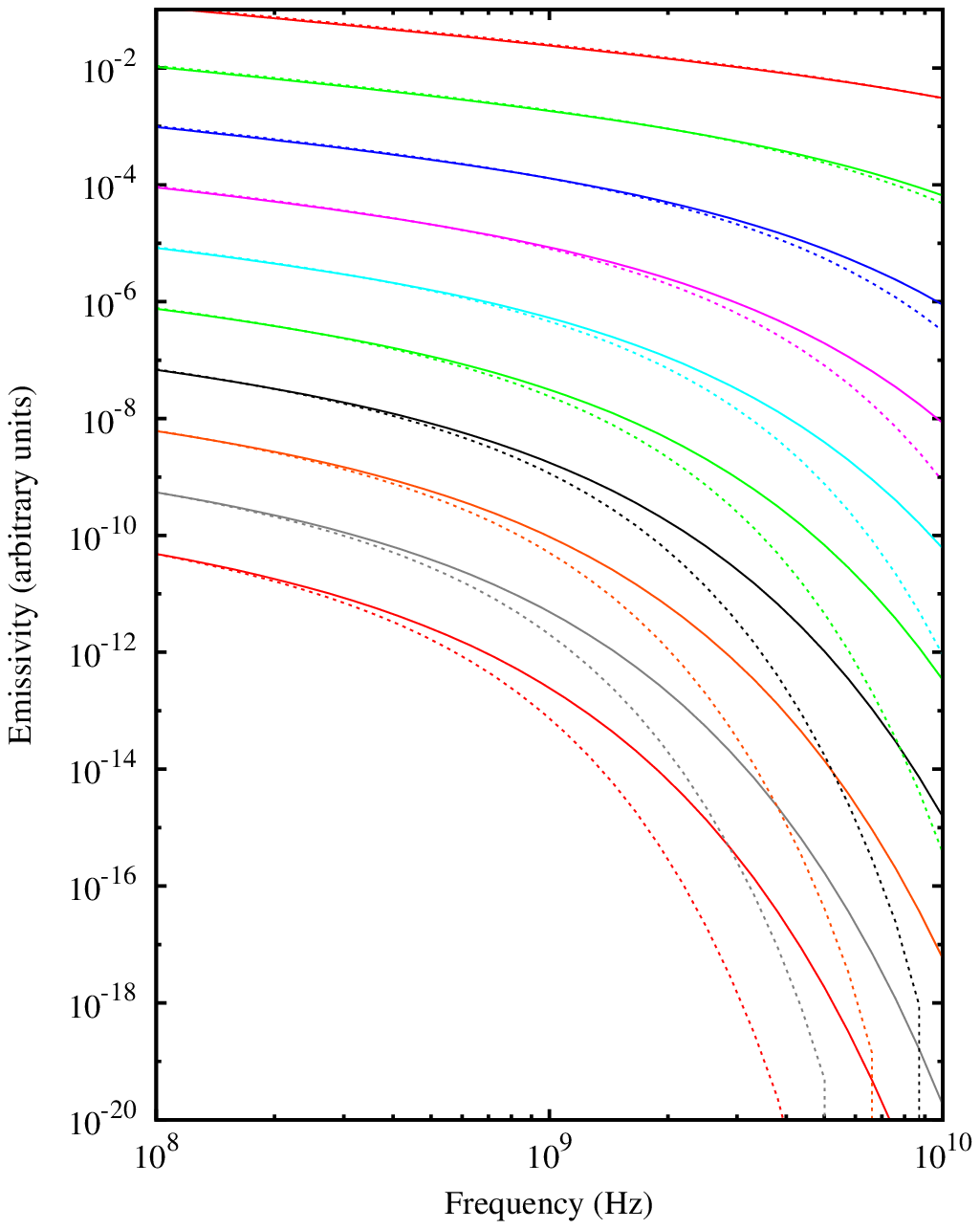}
\epsfxsize 8.7cm
\epsfbox{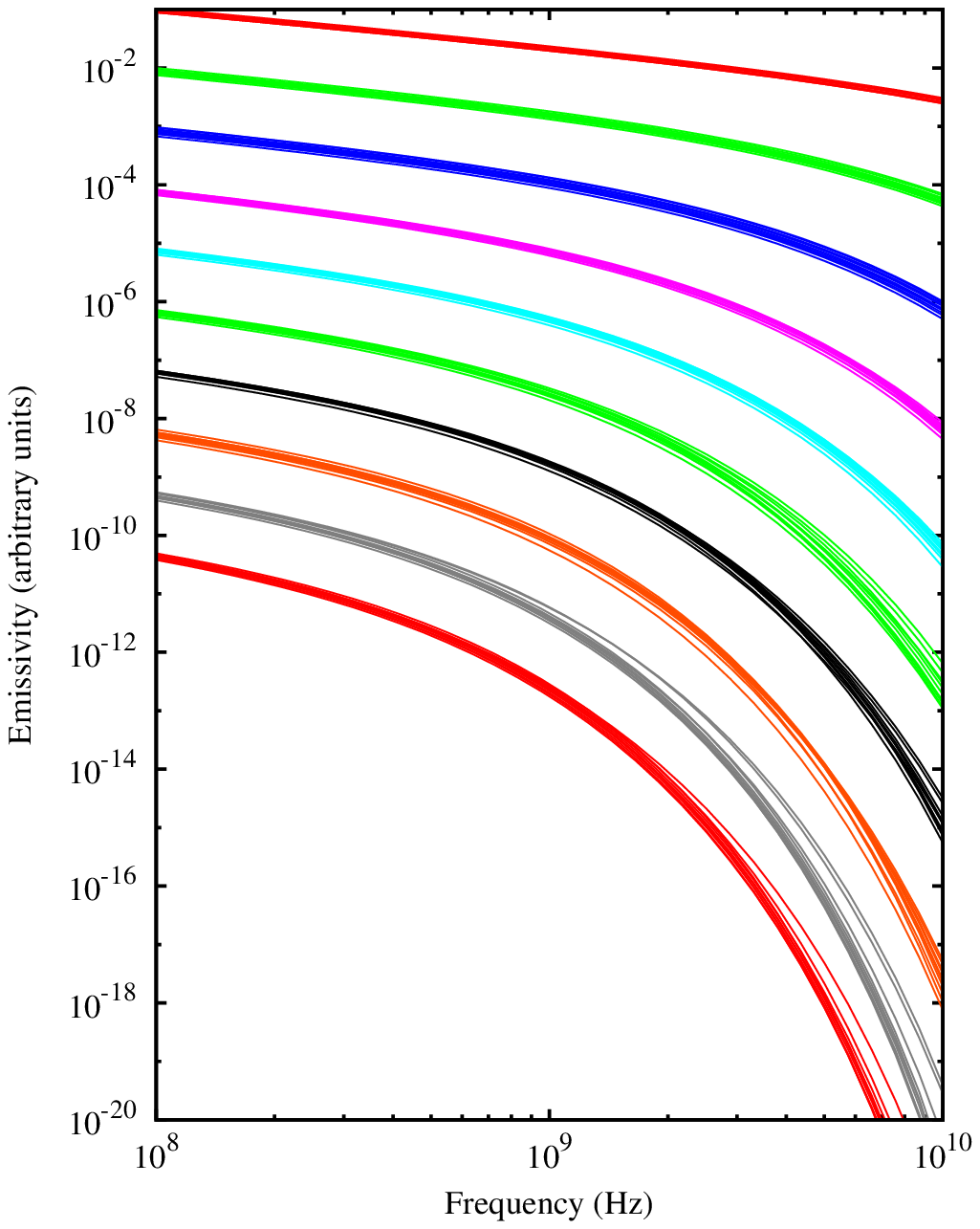}
\caption{Aged spectra in TJP and classical JP models. Left: solid
  lines show the the spectrum arising from numerically integrating the
  JP electron spectrum over the Maxwell-Boltzmann distribution of
  magnetic field strengths, with spectral ages of $10^7$ (flattest,
  top), $2 \times 10^7\dots$ $10^8$ years (steepest, bottom) with $B_0
  = 1$ nT. An order of magnitude in normalization is inserted between
  each curve. Dotted lines show the classical JP spectrum with a fixed
  magnetic field strength of 1 nT. Right: spectra from ten numerical
  realizations of finite-sized regions containing JP electron spectra,
  showing the dispersion that arises from the magnetic field structure
  in such regions, with each colour corresponding to runs for the age
  values plotted in the left-hand panel.}
\label{ages}
\end{figure*}

We may now consider spectral ageing. I adopt the JP electron spectrum,
i.e. 
\begin{equation}
N(E) = \begin{dcases}
   N_0 E^{-p} (1-ECt)^{(p-2)}&E<1/Ct\\
0&E\ge 1/Ct\\
\end{dcases}
\end{equation}
where $C$ is the constant such that
\begin{equation}
\frac{{\rm d}E}{{\rm d}t} = -CE^2
\end{equation}
i.e. $C$ takes account of both radiative and inverse-Compton losses:
\begin{equation}
C = \frac{4 \sigma_T}{3m_e^2c^3}\left(U_{\rm IC} + U_{B,\rm loss}\right)
\end{equation}
\citep{Longair10} where $U_{\rm IC}$ is the energy density in photons
and $U_{B,\rm loss} = B^2_{\rm loss}/2\mu_0$ is the energy density in
the effective field seen by the radiating
electrons over their post-acceleration lifetime, which in general will
be different from the local $B$.

The JP spectrum is the only one of the two `traditional' electron
spectra discussed above that it is reasonable to consider -- as
  discussed in Section 1, the
turbulent magnetic field that we impose would be expected to lead to
effective scattering as long as there is structure in the magnetic
field on scales comparable to the Larmor radius of the electrons. Such an electron energy
spectrum is expected in a non-uniform magnetic field model if, as
noted by \cite{Kaiser05}, the electron diffusion coefficient is
independent of energy -- in this case all electrons in a given volume
can be taken to have the same average loss history, and we can adopt
$B_{\rm loss} = B_0$. (This is also
true, independent of the diffusion coefficient, if inverse-Compton
losses dominate, i.e. $B_{\rm IC} \gg B_{\rm loss}$.)

This electron spectrum can then be used in eq. \ref{emissivity} to
compute the synchrotron emissivity (throughout this section of
  the paper we assume $s=0$). Fig.\ \ref{ages} shows plots of
the aged synchrotron spectrum for a Maxwell-Boltmann distribution of
magnetic field strength, assuming a uniform electron density that does
not depend on $B$ (for simplicity I refer to these as
Tribble-Jaffe-Perola, or TJP, spectra in what follows). To allow a
concrete comparison to real synchrotron spectra I set $B_0 = 1$ nT,
$B_{\rm loss} = 1$ nT, and $U_{\rm IC}$ appropriate for the $z=0$ CMB;
this is a regime appropriate for lobes of powerful radio galaxies in
the local Universe. The age $t$ is varied between $10^7$ and $10^8$
years. Also plotted, for comparison, are the JP synchrotron spectrum
appropriate for a single $B$-field value, $B = B_0$.

Several important points can be noted on this plot. First of all, we
see that the spectral shape is significantly altered by the adoption
of a Maxwell-Boltzmann distribution for $B$: the TJP spectrum cuts off
at higher frequencies and curvature is visible over a wider frequency
range. This is not at all surprising, since we can view these spectra
as the weighted sum of many individual single-field JP spectra, with
the break frequency for each being shifted according to
eq.\ \ref{nu-crit}. The spectra resemble the classical JP spectra in
the sense that, unlike the case in KP spectra, the spectral index
between any two frequencies $\nu_0$ and $\nu_1 = C\nu_0$ gets
monotonically steeper as $\nu_0$ moves upwards in frequency. However,
we also see that the spectra do not have the same shape as classical
JP spectra either -- a TJP spectrum matched to a single-frequency JP
spectrum at high and low frequencies is more strongly curved at
intermediate frequencies, and this excess curvature could in principle
be used to distinguish between them. Sensitive observations spanning a
broad frequency range, such as will soon be provided by combining
LOFAR and JVLA data, will be needed to see if these models can be
tested.

Secondly, we
see that the normalization of the two spectra is very similar; this
follows from the type of argument we used above to discuss power laws. 
Since the TJP extends to higher frequencies, we
would expect its normalization to be slightly below that of the
uniform-field spectrum, which is observed.

Finally, we see that in the realizations of the TJP spectra using the
numerically modelled regions, plotted in Fig.\ \ref{ages} (right-hand
panel), there is very substantial scatter, particularly in
steep-spectrum regions. This comes about because the magnetic field
structure that we have generated has its maximum power on scales
comparable to that of the box itself -- they are not simply a
resolution effect, but show the real expected degree of scatter in a
Gaussian random magnetic field with a power spectrum of the type we
have chosen. We expect this to be a realistic description of
  regions much smaller than the lobe size, where the largest scale of
  surface brightness variation is larger than the region size, so long
  as the power spectral index $\zeta$ is not too flat. As the plots
show, finite-size realizations (which is what we always see in
practice) may differ significantly from the idealized spectra derived
by integration of eq. \ref{emissivity}. In steep-spectrum regions of
highly aged spectra the dispersion in the predicted flux density can
easily be an order of magnitude (and would be larger still for
  $s>0$).

As pointed out by \cite{Tribble94}, the emissivity for such an aged
electron spectrum, for appropriately high frequencies, is a very
strong function of magnetic field strength, because the local $B$
determines the break frequency of the local synchrotron emissivity. It
is a general prediction of models in which the observed filamentary
structure is determined by magnetic field variation that the
filamentary structures will become more prominent as we move to higher
observing frequencies, or, equivalently, that brighter regions will
have flatter spectra than fainter regions.

If real radio galaxies have spectra described by the TJP model, what
are the consequences of the widespread fitting of the standard JP
model with its assumption of a uniform magnetic field? One is obvious:
for a given electron age, the TJP spectra cut off at higher
frequencies, so we would expect JP models to underestimate the true
spectral age. Although this effect is not large (Fig.\ \ref{ages}) it
is in the sense expected from the generally accepted conclusion
that the spectral ages underestimate the dynamical ages of radio
galaxies \citep[e.g.][]{Eilek96}. Another consequence is slightly less
obvious. Because the TJP models are more curved than the JP models at
low frequencies, fitting a JP model over a finite frequency range,
with the `injection index' -- i.e. the low-frequency spectral index --
left as a free parameter, will tend to produce artificially high
injection indices -- the TJP model flattens more at frequencies below
those that are observed than the JP model expects. The magnitude of
the effect will depend on the frequencies used and the location of the
spectral break with respect to them, and so is hard to quantify, but
some simple tests suggest that the effect is not negligible. Users of
the JP model should treat best-fitting injection indices with caution;
however, see Harwood \etal (submitted), for an example where the TJP
and JP models do not give significantly different answers.

\section{Inverse-Compton measurements}

\begin{figure*}
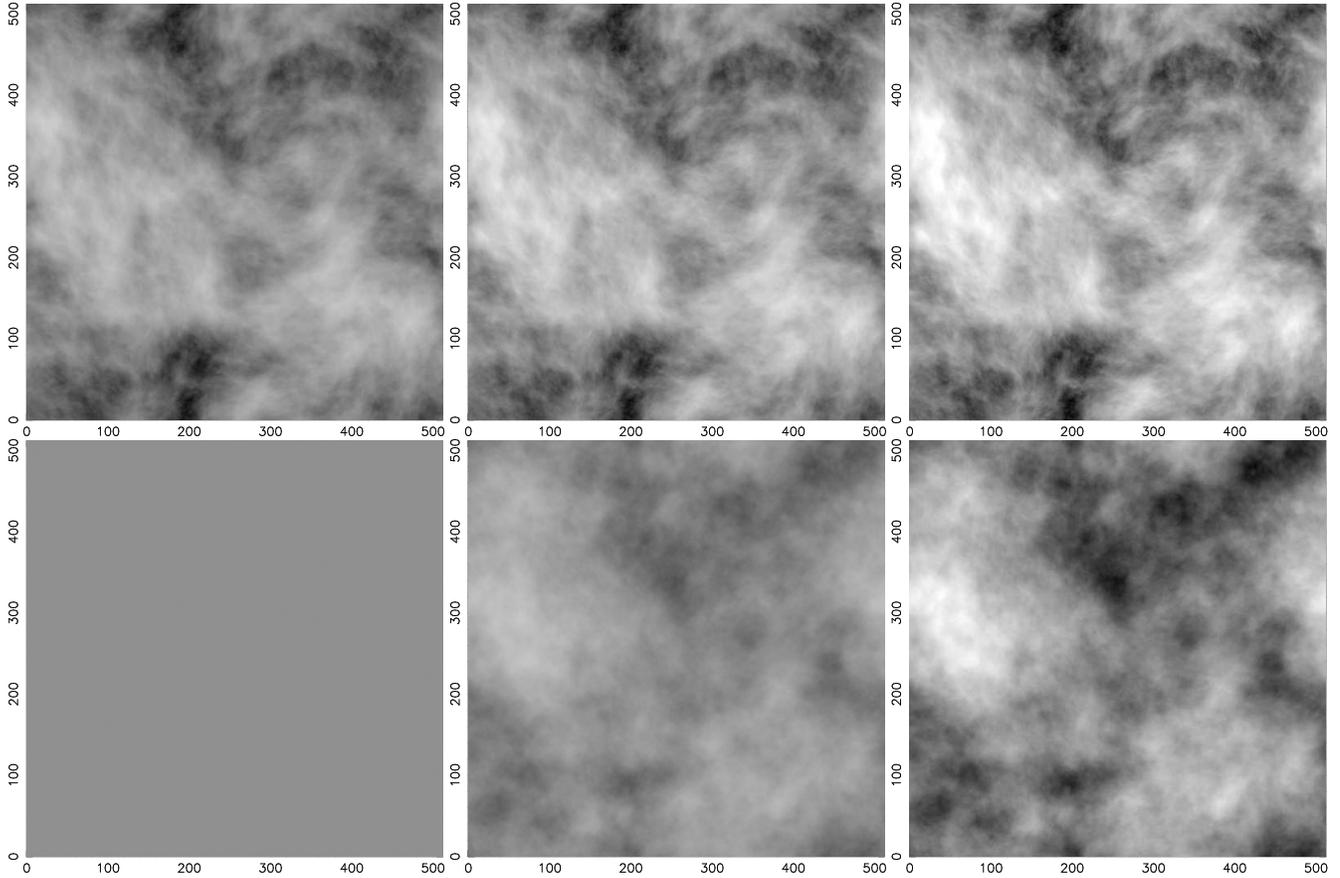

\hbox{\epsfxsize 5.8cm
\epsfbox{synch-0.ps}
\epsfxsize 5.8cm
\epsfbox{synch-1.ps}
\epsfxsize 5.8cm
\epsfbox{synch-2.ps}
}
\hbox{\epsfxsize 5.8cm
\epsfbox{ic-0.ps}
\epsfxsize 5.8cm
\epsfbox{ic-1.ps}
\epsfxsize 5.8cm
\epsfbox{ic-2.ps}
}
\caption{(Top) synchrotron and (bottom) inverse-Compton emission from
  the same magnetic field distribution, with the electron energy
  density dependence on the magnetic field strength increasing from
  left to right: the three panels show $s=0$, 1 and 2 respectively.
  Grey levels are logarithmic and are the same for each set of images.}
\label{inverse-compton}
\end{figure*}

\subsection{Scattering of the CMB}

If the electron number density is constant, the results of section
\ref{powerlaw} are good news for inverse-Compton estimates of magnetic
field strength that involve scattering of an isotropic photon field
such as the CMB. The emissivity from this process depends on the
number density of low-energy electrons (e.g. $\gamma \sim 1000$ for
scattering of CMB photons to keV X-ray energies) and therefore on
the normalization of the electron energy spectrum, $N_0$. The magnetic
field strength is then essentially estimated from synchrotron
observations using
eq.\ \ref{emiss-approx} (in the case of a power law), so
\begin{equation}
B \propto \left(\frac{J(\nu)}{N_0}\right)^{\frac{2}{p+1}}
\end{equation}
where the assumption is that there is a single value of $B$ throughout
the region of interest.

As we saw in Section \ref{powerlaw}, replacing a single magnetic field
strength $B$ with a distribution of field strengths characterized by a
mean squared field $B_0^2$ barely changes the normalization of the
relationship between $N_0$ and synchrotron emissivity, $J(\nu)$
(eq.\ \ref{normchange}). Therefore magnetic field estimates derived in
this way are good estimates of the mean field $B_0$. Even if
we consider the case where the electron energy density scales with the
field energy density ($s=2$ in eq. \ref{s-definition}) we saw that the
normalization of the $J(\nu)$ -- $N_0$ relationship changes at most by
a factor 1.5 (in the sense that more emission is produced for a given
mean number density of electrons). The integrated inverse-Compton emissivity,
which depends on the mean value of $N_0$, is unchanged in such models,
and so the change in synchrotron emissivity would mean that the mean magnetic
field strength would be overestimated by at most a factor $\sim 1.3$
(for $p=2$). Similar results will hold for more complex spectra, such
as the curved spectra of Section \ref{aged}, though here additional
systematic errors in the standard calculations will result from
incorrect assumptions about the mapping between the electron energy
spectrum and the synchrotron spectrum.

One interesting feature of inverse-Compton emission is that it is
possible to test the class of models where the electron energy/number
density scales with field strength (i.e. where $s>0$ in eq.
\ref{s-definition}); as noted above, these models have little effect
on global properties of the synchrotron spectrum such as the power
spectrum index. If $s>0$, we expect to see structure in the
inverse-Compton emission that will correlate with structure in the
synchrotron emission -- the correlation will not be perfect (since the
emissivities in inverse-Compton and synchrotron are always different
non-linear functions of the local magnetic field strength) but will be
more apparent for high values of $s$. We can visualize this reasonably
well, without doing a detailed calculation by assuming that the
inverse-Compton emissivity scales as $N_0$; examples are given in
Fig.\ \ref{inverse-compton}. Sources that show bright, resolved
inverse-Compton emission allow us to put constraints on the value of
$s$ (Goodger \etal, in prep.).

\subsection{Synchrotron self-Compton}

Synchrotron self-Compton (SSC) emission is a much more challenging
problem, since the photon field is no longer isotropic and the
spectrum from every illuminating point may be different. We expect the
SSC emission from a region with variable magnetic field strength to be
non-uniform even if the electron density is uniform. To calculate the
SSC from our numerical models we need, in principle, to consider the
illumination of every point by every other point, integrating the full
anisotropic inverse-Compton kernel over the distribution of electrons
and incoming photons. This is computationally very expensive, and so
shortcuts of various sorts are normally taken to avoid having to do
the full integration
\citep[e.g.,][]{Hardcastle+02,Hardcastle+Croston11}. In order to get
an estimate of the effects of inhomogeneous magnetic fields, I
consider only power-law spectra, which has the effect that we need not
consider a different synchrotron spectrum from each volume element,
though I still take account of the anisotropic emissivity of
synchrotron radiation. I treat anisotropic inverse-Compton scattering
using the approach of \cite{Hardcastle+02}, based on the
inverse-Compton formulations of \cite{Brunetti00}, which are
integrated over the power-law photon and electron distributions,
assuming scattering from a fixed radio frequency to 1-keV X-rays (i.e.
well away from the Klein-Nishina regime). All of these simplifications
reduce the computation for each pair of volume elements to a few
simple operations, rather than any numerical integration. Even so, the
$N^6$ dependence of execution time makes the use of $512^3$ boxes, as
in earlier parts of the paper, impossible, and the results in what
follows are based on $100^3$ boxes, which can be run in a few minutes
on a moderate number of compute cores. (For consistency, I continue to
use cubical boxes even though the geometry affects the emissivity for
the SSC process, because of edge effects, and a cube is not a
realistic shape for an astrophysical object -- the scaling results
should be the same whatever the shape we consider.)

\begin{figure*}
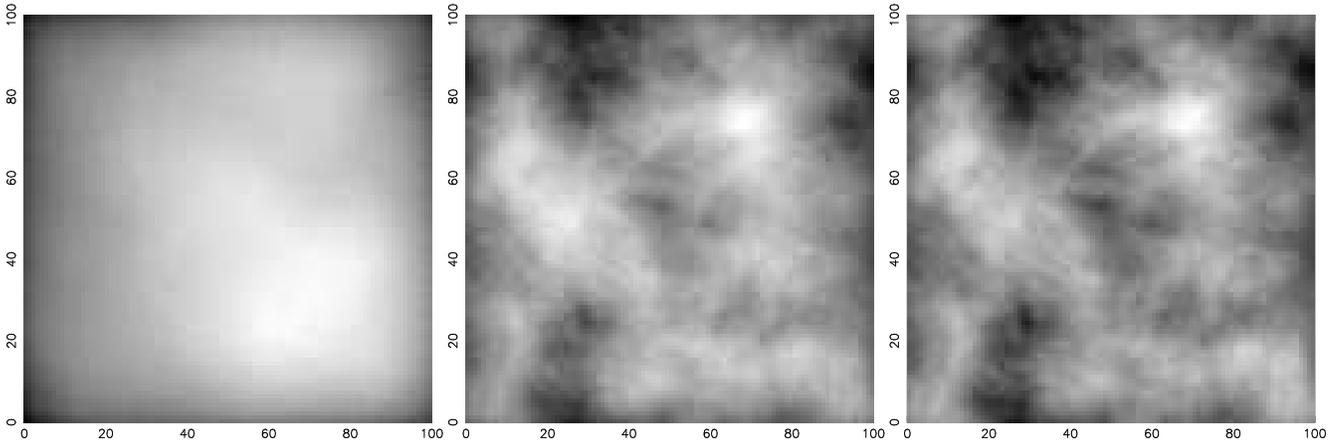

\hbox{\epsfxsize 5.8cm
\epsfbox{out-100-0.ps}
\epsfxsize 5.8cm
\epsfbox{out-100-1.ps}
\epsfxsize 5.8cm
\epsfbox{out-100-2.ps}
}
\caption{SSC emission from a $100^3$ simulated box, on the assumption
  of, from left to right, $s=0$, $s=1$ and $s=2$; all three boxes have
  the same magnetic field distribution. The dynamic range is allowed
  to vary (i.e. the greylevels are not matched) so that structure in
  the $s=0$ image can be seen. The dynamic range of the 3 images
    (the ratio of surface brightness represented by white points to
    those represented by black points) is respectively 3.2, 7.4 and 14.5.}
\label{ssc}
\end{figure*}

\begin{figure}
\epsfxsize 8.5cm
\epsfbox{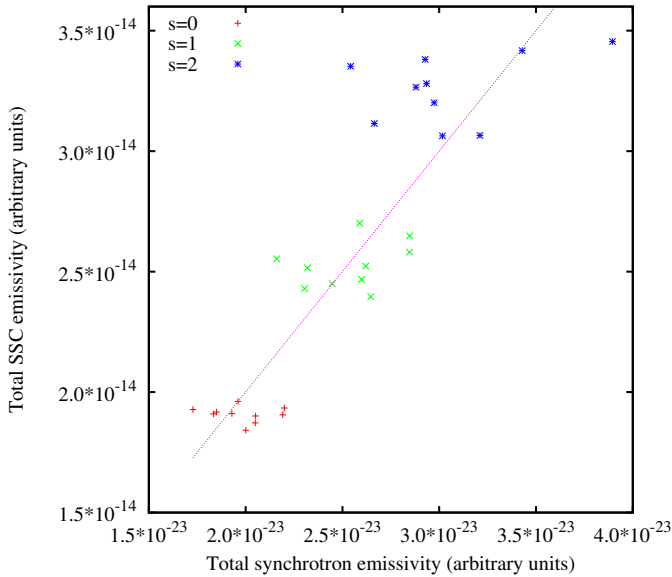}
\caption{Total synchrotron and SSC emissivity for ten realizations of
  $100^3$ boxes with $s=0$, 1 and 2. Both SSC and synchrotron emission
  increase as $s$ increases, and the scaling is very nearly linear
  (for comparison, the purple line shows a linear scaling normalized
  by eye to pass through the data points). The scatter in the
  synchrotron emissivity is greater than that in SSC.}
\label{s-ssc}
\end{figure}

We would expect that the mean SSC emissivity from a box where $s=0$,
so that the electron density is uniform, would be comparable to the
emissivity for the case where both the electrons and field (and hence
the synchrotron emissivity) are uniform, and this is indeed what we
see: the SSC emissivity is not uniform in finite-sized realizations,
because of the non-uniform synchrotron emissivity, but it is much less
strongly structured than the synchrotron emissivity because photons
from the whole volume illuminate every region. For $s>0$, the results
are expected to be more interesting, because now higher densities of
electrons are concentrated in regions of higher photon energy density,
and indeed we see that structure begins to appear in the
inverse-Compton images (Fig.\ \ref{ssc}). The mean inverse-Compton
emissivity also increases, by a factor 1.3 for $s=1$ and 1.6 for $s=2$
-- thus it tracks, rather closely, the increase in synchrotron
emissivity discussed in Section \ref{powerlaw}, as we might expect
(Fig.\ \ref{s-ssc}). The effect of such models would therefore be on
average to increase both synchrotron and SSC emissivity, more or less
in step. Since for power-law SSC we know that
\begin{equation}
J_{\rm SSC} \propto N_0 J_{\rm synch}
\end{equation}
and so from eq.\ \ref{emiss-approx}
\begin{equation}
B \propto \left(\frac{J_{\rm synch}^2}{J_{\rm SSC}}\right)^\frac{2}{p+1}
\end{equation}
then the fact that $J_{\rm SSC}$ averaged over the volume scales
roughly as $J_{\rm synch}$ means that, as with the CMB
inverse-Compton, $B$-field estimates derived in the conventional way
will be wrong only by a factor $\sim 1.3$ for $p=2$, $s=2$. In
finite-sized regions there will of course be scatter in both the
synchrotron and SSC emission, but the former is dominant (Fig.
\ref{s-ssc}). If the spectra have substantial curvature, as in the
models of Section \ref{aged}, then the dynamic range of the SSC images
would be expected to increase significantly (particularly as
high-frequency photons in the synchrotron emission are important for
efficient SSC scattering to X-rays by low-energy electrons). We might
also expect the SSC results to be somewhat dependent on the power
spectrum of fluctuations, $\zeta$.

As with the CMB inverse-Compton, we could in principle test for $s>0$
by looking at the strength of the correlation between SSC and
synchrotron surface brightness. However, most well-resolved lobes are
unlikely to be dominated by SSC. One exception is Cygnus A
\citep{Hardcastle+Croston11} but X-ray inverse-Compton emission is only
barely detected in this source against the thermal fore/background
emission from the host cluster, and mapping it spatially will almost
certainly require next-generation instruments.

\section{Summary and conclusions}

The key points from the analysis presented in this paper are:
\begin{enumerate}
\item I have given an explicit formula (eq.\ \ref{emissivity}) for the
  synchrotron emissivity in a large region with a Gaussian random magnetic field
  -- this was implicit, but never written out in this simple form, in
  the earlier work of Tribble.
\item A comparison of the synchrotron emissivity between uniform-field
  models and more realistic random-field models is particularly
  simple if we assume that the two have the same mean magnetic field
  energy density: in this case we find that the mean emissivity is
  almost unchanged, a result that can be derived either from simple
  physical arguments or from integrating the synchrotron emissivity
  function over $B$. If the electron energy density scales as some
  power of the magnetic field strength, while retaining the same mean
  value, then the mean emissivity increases, but not by large factors.
  Standard minimum-energy arguments that rely on uniform-field
  assumptions for $B$ are not wrong by large factors.
\item JP electron spectra in random fields give rise to a
  well-defined synchrotron spectrum (the `TJP spectrum') that is significantly different
  from the standard JP synchrotron spectrum; I have presented a simple
  recipe for calculating this model numerically. The applicability of
  the TJP spectrum to real sources can
  be tested using fits to broad-band data from the current generation
  of radio telescopes. Numerical modelling shows that there may be
  very significant scatter in finite-sized realizations of this
  spectrum due to structure in the synchrotron emissivity.
  Fitting a TJP spectrum with a JP model may cause the low-frequency
  spectral index (`injection index') to be overestimated and the
  spectral ages to be underestimated.
\item Random fields give rise to inverse-Compton emissivity which is
  generally very close to the expectation from uniform-field models:
  the energy densities in magnetic field estimated with the standard
  assumptions from inverse-Compton observations are therefore good
  estimates of the mean energy density, if the electron energy density
  is independent of field strength. If the electron energy density
  scales as some power $s$ of the magnetic field strength, the field
  strength will be overestimated, but only by small factors for
  plausible values of $s$. The relationship between the resolved
  surface brightness of synchrotron and inverse-Compton emission
  depends on the value of $s$ and this can in principle be tested by
  observations.
\end{enumerate}

Harwood \etal\ (submitted) discuss fits of the TJP spectrum
to real data, while Goodger \etal\ (in prep.) will attempt to
constrain $s$ from observations of resolved inverse-Compton emission.
Applications of these models to the polarization of radio galaxies and
their depolarization by the external medium will be discussed in a
later paper.

\section*{Acknowledgements}

This work has made use of the University of Hertfordshire Science and
Technology Research Institute high-performance computing facility. 
I thank Judith Croston for helpful comments on a draft of the paper,
and an anonymous referee for a constructive report which has
  helped to improve the paper.

\bibliographystyle{mn2e}
\renewcommand{\refname}{REFERENCES}
\setlength{\bibhang}{2.0em}
\setlength\labelwidth{0.0em}
\bibliography{../bib/mjh,../bib/cards}

\end{document}